\documentclass[aps,prl,twocolumn,floatfix,showpacs,superscriptaddress]{revtex4-1}
\usepackage{amssymb}

\usepackage{epsfig}
\usepackage{amsmath}

\begin{document}

\title{The origin of electron-hole asymmetry in graphite}

\author{P. \surname{Plochocka}}
\affiliation{Laboratoire National des Champs Magn\'etiques Intenses, CNRS-UJF-UPS-INSA, 143 avenue de Rangueil, 31400
Toulouse, France}\affiliation{Laboratoire National des Champs Magn\'etiques Intenses, CNRS-UJF-UPS-INSA, 25 avenue des
Martyrs, 38042 Grenoble, France}
\author{P. Y. \surname{Solane}}
\affiliation{Laboratoire National des Champs Magn\'etiques Intenses, CNRS-UJF-UPS-INSA, 143 avenue de Rangueil, 31400
Toulouse, France}
\author{R. J. \surname{Nicholas}}
\affiliation{Clarendon Laboratory, Physics Department, Oxford University, Parks Road, Oxford OX1 3PU, United Kingdom}
\author{J. M. \surname{Schneider}}
\author{B. A. \surname{Piot}}
\author{D. K. \surname{Maude}}
\affiliation{Laboratoire National des Champs Magn\'etiques Intenses, CNRS-UJF-UPS-INSA, 25 avenue des Martyrs, 38042
Grenoble, France}
\author{O. \surname{Portugall}}
 \affiliation{Laboratoire National des Champs Magn\'etiques Intenses, CNRS-UJF-UPS-INSA, 143 avenue de
Rangueil, 31400 Toulouse, France}
\author{G. L. J. A. \surname{Rikken}}
 \affiliation{Laboratoire National des Champs Magn\'etiques Intenses, CNRS-UJF-UPS-INSA, 143 avenue de
Rangueil, 31400 Toulouse, France}\affiliation{Laboratoire National des Champs Magn\'etiques Intenses,
CNRS-UJF-UPS-INSA, 25 avenue des Martyrs, 38042 Grenoble, France}

\date{\today}

\begin{abstract}
The electron hole asymmetry has been measured in natural graphite using magneto-optical absorption measurements. A
splitting is observed for the transitions at both the $K$-point and the $H$-point of the Brillouin zone of graphite
where the effect of trigonal warping vanishes. This result is fully consistent with the SWM Hamiltonian providing the
free electron kinetic energy terms are taken into account. An identical electron-hole asymmetry should be present in
graphene.
\end{abstract}


\maketitle

The band structure of graphite has been calculated by Slonczewski and Weiss (SW) in the late
fifties~\cite{Slonczewski58}. Based upon detailed group theoretical considerations the SW Hamiltonian, with its seven
tight binding parameters $\gamma_0,..,\gamma_5, \Delta$, can be exactly diagonalized to give the band structure. Due to
the inter layer coupling the in-plane dispersion depends on the momentum $k_z$ parallel to the $c$-axis. McClure
derived the magnetic Hamiltonian for the case when the magnetic field is applied parallel to the
$c$-axis~\cite{McClure60}. The so called Slonczewski, Weiss and McClure (SWM) Hamiltonian has infinite order since the
trigonal warping term $\gamma_3$ couples Landau levels with orbital quantum number $n$ to Landau levels with quantum
number $n+3$. This coupling breaks the dipole selection rule and gives rise to a large number of harmonics in the
cyclotron resonance. Nakao showed that the infinite Hamiltonian can be successfully truncated to a reasonable size and
numerically diagonalized to find the eigen values~\cite{Nakao76}. At the $H$-point the effect of $\gamma_3$ vanishes
and the SWM Hamiltonian can be analytically solved to give a Landau level energy spectrum which depends only on
$\gamma_0$. This is the origin of a widespread misconception in the literature, including our own work, that there is
no electron-hole asymmetry at the $H$-point.

The electronic properties of graphite are well documented in the
literature~\cite{Soule64,Williamson65,Schroeder68,Woollam70,Woollam71a,Shimamoto98,Lukyanchuk04,Lukyanchuk06,Mikitik06,Schneider09,Zhu09,Schneider10,Lukyanchuk10,Schneider10a,Hubbard11}.
In particular, magneto-optical techniques have been extensively used to probe the Landau level energy spectrum at the
$H$ and $K$-points where there is a joint maximum in the optical density of
states~\cite{Doezema79,Orlita08,Orlita09,Chuang09,Ubrig11,Tung11}. This data was for the most part analyzed using the
effective bi-layer model~\cite{Koshino08} for graphite with only two parameters, $\gamma_0$ and an effective inter
layer coupling $2\gamma_1$. The splitting of the $K$-point transitions in the magneto-reflectance data was analyzed
within the effective bi-layer model by including electron-hole asymmetry due to the non vertical coupling term
$\gamma_4$ phenomenologically~\cite{Chuang09}. In our previous work~\cite{Ubrig11} the observed splitting of the
$H$-point transitions was not assigned to electron-hole asymmetry as there is no trigonal warping at the $H$-point, so
the effect of $\gamma_4$ vanishes.

In this letter we show that electron-hole asymmetry exists for all values of $k_z$ and is an inherent part of the SWM
Hamiltonian through the often neglected free electron kinetic energy terms. The asymmetry should lead to an observable
splitting of both the $H$ and $K$-point optical transitions. Extending our previous magneto-optical on natural graphite
to lower energies, lower temperatures, and higher magnetic fields we show that a splitting of both the $H$ and
$K$-point transitions due to the electron-hole asymmetry is observed. The size of the splitting at the $H$-point is in
good agreement with the predicted electron-hole asymmetry. The splitting of the $K$-point transitions is also found to
be dominated by the free electron terms with $\gamma_4$ and $\gamma_5$ playing only a secondary role.

Nakao~\cite{Nakao76} derived an explicit form for Landau level energy spectrum at the $H$-point. Unfortunately, when
writing the expression Nakao neglected for simplicity the small free electron kinetic energy terms $\hbar^2 k^2/2m$,
where $k$ is the in plane wave vector and $m$ is the \emph{free} electron mass. The free electron terms are quantized
in a magnetic field and their values are significant for all magnetic fields. The SWM Hamiltonian can easily be
diagonalized at the $H$-point and the correct expression for the Landau level spectrum, including the free electron
terms is,

\begin{equation}\label{EqHpoint1}
E_{3\pm}^n = \frac{\Delta \pm \sqrt{(\Delta+\hbar^2 s/2 m)^2 + 3 n s \gamma_{0}^2 a_{0}^2}}{2} + \frac{n \hbar^2 s}{2
m},
\end{equation}
\begin{equation}
E_{1,2}^n = \frac{\Delta \pm \sqrt{(\Delta-\hbar^2 s/2 m)^2 + 3 (n+1) s \gamma_{0}^2 a_{0}^2}}{2} + \frac{(n+1) \hbar^2
s}{2 m}\nonumber,
\end{equation}
where $n=0,1,2,...$ is the orbital quantum number, $s=2 e B / \hbar$ and $a_{0}=0.246$~nm. The Zeeman term has been
omitted since it simply shifts the energies by $\pm g \mu_B B / 2$ and can easily be added if required. At the
$H$-point the electron hole asymmetry is provided by the free electron term $n\hbar^2 s/2m$. Thus the dipole allowed
transitions, $E_{3-}^n \rightarrow E_{3+}^{n+1}$ and $E_{3-}^{n+1} \rightarrow E_{3+}^{n}$ will be split by $\delta E =
\hbar^2 s/2m \simeq 0.23$~meV/T. Note that $\hbar^2 s/2 m \ll s\gamma_{0}^2 a_{0}^2$  so that to a very good
approximation $E_{3_\pm}^{n+1}=E_{1,2}^{n}$ \emph{i.e.} the Landau ladders remain doubly degenerate at the $H$-point.
Note that the free electron term has the expected phase (0) for massless Dirac fermions.

In a similar way, the bi-layer expression~\cite{Koshino08} can be modified phenomenologically to include the free
electron term for massive fermions with a phase of (1/2) at the $K$-point
\begin{multline}\label{Bilayer}
E^{n}_{3\pm}=\pm\frac{1}{\sqrt{2}}\left[(\lambda\gamma_1)^2+(2n+1)\varepsilon^2\right.\\
\left.-\sqrt{(\lambda\gamma_1)^4 +2(2n+1)\varepsilon^2(\lambda\gamma_1)^2+\varepsilon^4}\right]^{1/2}\\
+ \frac{(n+\frac{1}{2}) \hbar^2 s}{2 m},
\end{multline}
where $n=0,1,2...$, $\lambda=2$, $\varepsilon=v_f\sqrt{2e\hbar B}$ is the characteristic magnetic energy, $v_f=\sqrt3 e
a_0 \gamma_0/2\hbar$ is the Fermi velocity. Eq(\ref{Bilayer}) has not been derived explicitly, however, we have
verified that the predicted behavior is in exact agreement with the SWM calculation with $\gamma_3,..,\gamma_5,
\Delta=0$. The full SWM model has quantum numbers $-1,0,1,2,...$ and there are two special Landau levels (LL0 and LL-1)
whose energy remains close to zero. The bilayer model (Eq.(\ref{Bilayer})) correctly predicted the energy of LL-1 using
$n=0$. LL0 is missing but can be reproduced accurately between $0-150$~T using $n=0$ if the free electron term is
replaced by $(n+3/2) \hbar^2 s / 2 m - 16 (\hbar^2 s / 2 m)^2$.

\begin{figure}[]
\begin{center}
\includegraphics[width= 7.5cm]{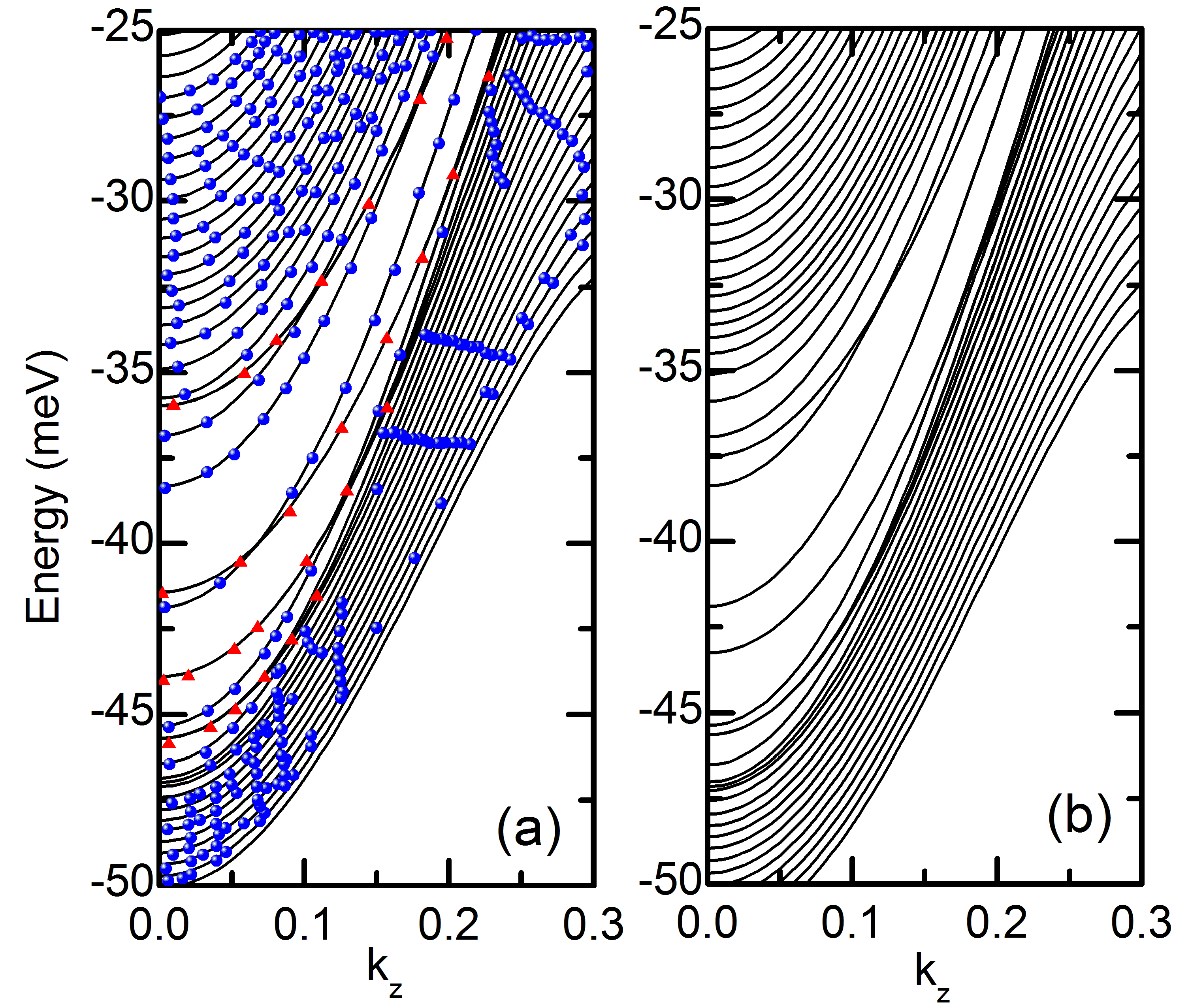}
\end{center}
\caption{(Color online) (a) Calculated Landau level dispersion (solid lines) along $k_z$ using the SWM parameters of
Nakao~\cite{Nakao76} and including the free electron terms. For comparison the calculated values of Nakao (symbols) are
shown. (b) Calculated Landau level dispersion along $k_z$ neglecting the free electron terms.}\label{fig1}
\end{figure}

Before presenting the experimental data, the importance of the free electron kinetic energy terms is demonstrated by
numerically diagonalizing the truncated $600 \times 600$ SWM matrix for a magnetic field $B=0.3$~T using the SWM
parameters of Nakao~\cite{Nakao76} to allow a comparison. The calculated Landau level dispersion along $k_z$ is shown
in Fig.\ref{fig1}(a) including the free electron terms. The symbols (circles and triangles) in Fig.\ref{fig1}(a) are
taken from the calculations of Nakao at the same magnetic field (Fig.~3 of Ref.\cite{Nakao76}). The triangles
distinguish the triply degenerate Landau levels, which have a markedly different dispersion along $k_z$ and correspond
to leg orbits. Clearly there is perfect agreement between the two calculations. On the other hand, the calculations in
Fig.\ref{fig1}(b) which neglect the free electron terms are significantly different. Notably, the electron cyclotron
energy is underestimated, while the hole cyclotron energy is overestimated. Thus, the free electron terms have to be
included in the SWM Hamiltonian if the correct energy spectrum is to be obtained. As our SWM calculations agree
perfectly with the results of Nakao, we conclude that the free electron terms were omitted from Eq(9) of
Ref.\cite{Nakao76}, but included in the numerical calculations of Nakao.

For the magneto-transmission measurements suitably thin samples were fabricated by exfoliating natural graphite. The
measurements were performed in pulsed fields $\leq 60$~T ($\simeq 400$~mS). A tungsten halogen lamp provides a broad
spectrum in the visible and near infra-red range and the absorption is measured in the Faraday configuration with the
$c$-axis of the graphite sample parallel to magnetic field.  A nitrogen cooled InGaAs photodiode array, or an extended
InGaAs detector analyzed the transmitted light dispersed by a spectrometer. The use of two detectors allows us to cover
a wide energy range $0.6-1.1$~eV. Differential transmission spectra were produced by normalizing all the acquired
spectra by the zero field transmission. Measurements to higher fields $\leq 150$~T were performed using a
semi-destructive technique and pulse lengths of $\simeq 10$~$\mu$s and the transmission of a polarized CO laser
($0.229$~eV) measured as a function of the magnetic field using a nitrogen cooled HgCdTe photodiode coupled with a 200
MHz low noise amplifier and an infrared tunable wave plate.

\begin{figure}[]
\begin{center}
\includegraphics[width= 7.5cm]{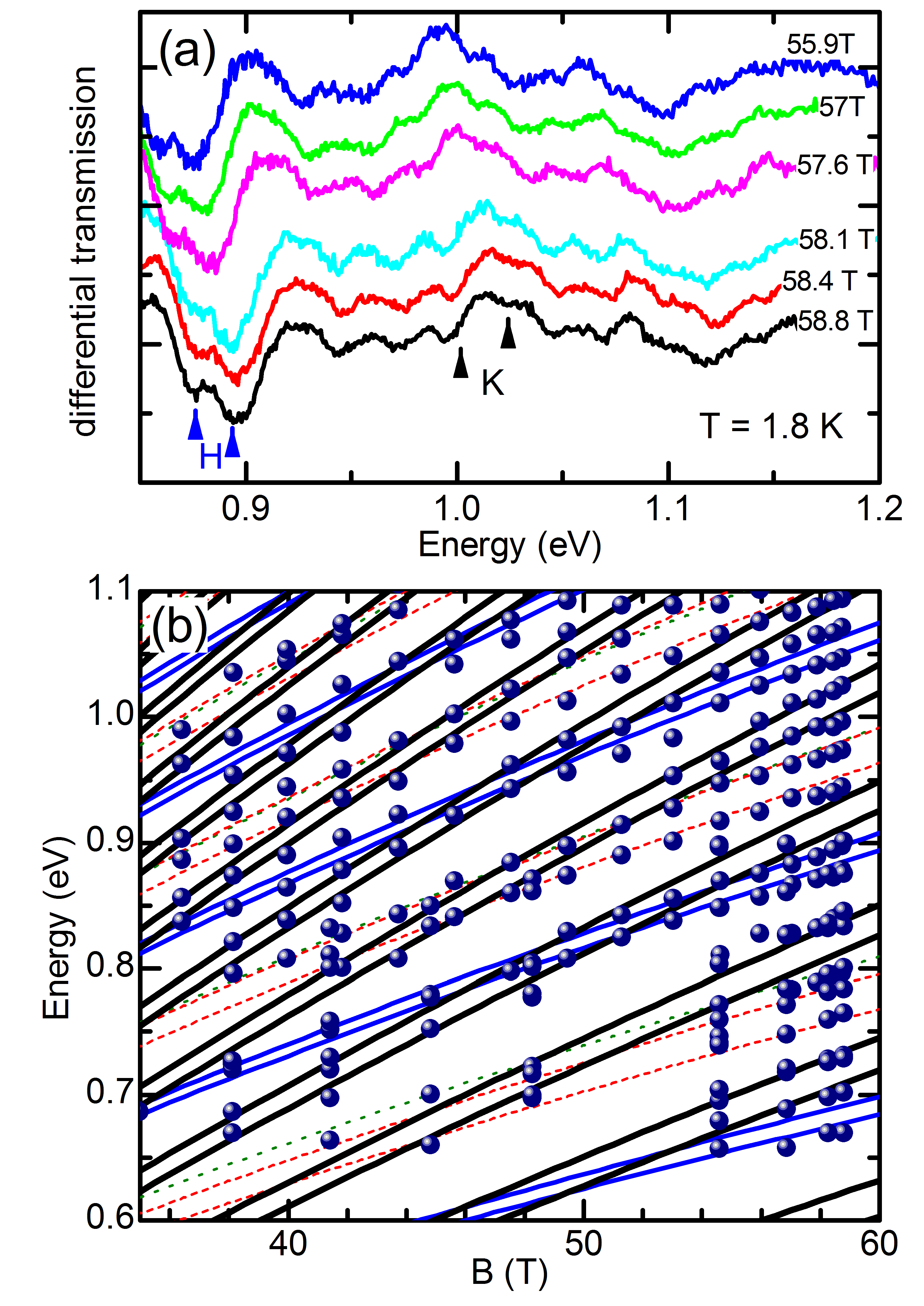}
\end{center}
\caption{(Color online) (a) Differential magneto-transmission spectra of natural graphite measured at magnetic fields
in the range $55-59$~T at $T\simeq 1.8$~K. (b) Magnetic field dependence of the observed optical transitions in natural
graphite. The calculated SWM energies of the transitions are shown as lines: $H$-point $\Delta n=\pm 1$ (thin blue
lines), ``effective'' $H$-point $\Delta n=\pm 2$ (dashed red lines), $\Delta n=0$ (dotted green lines) and $K$ point
$\Delta n=\pm 1$ (thick black lines).} \label{fig2}
\end{figure}

\begin{figure}[]
\begin{center}
\includegraphics[width= 7.0cm]{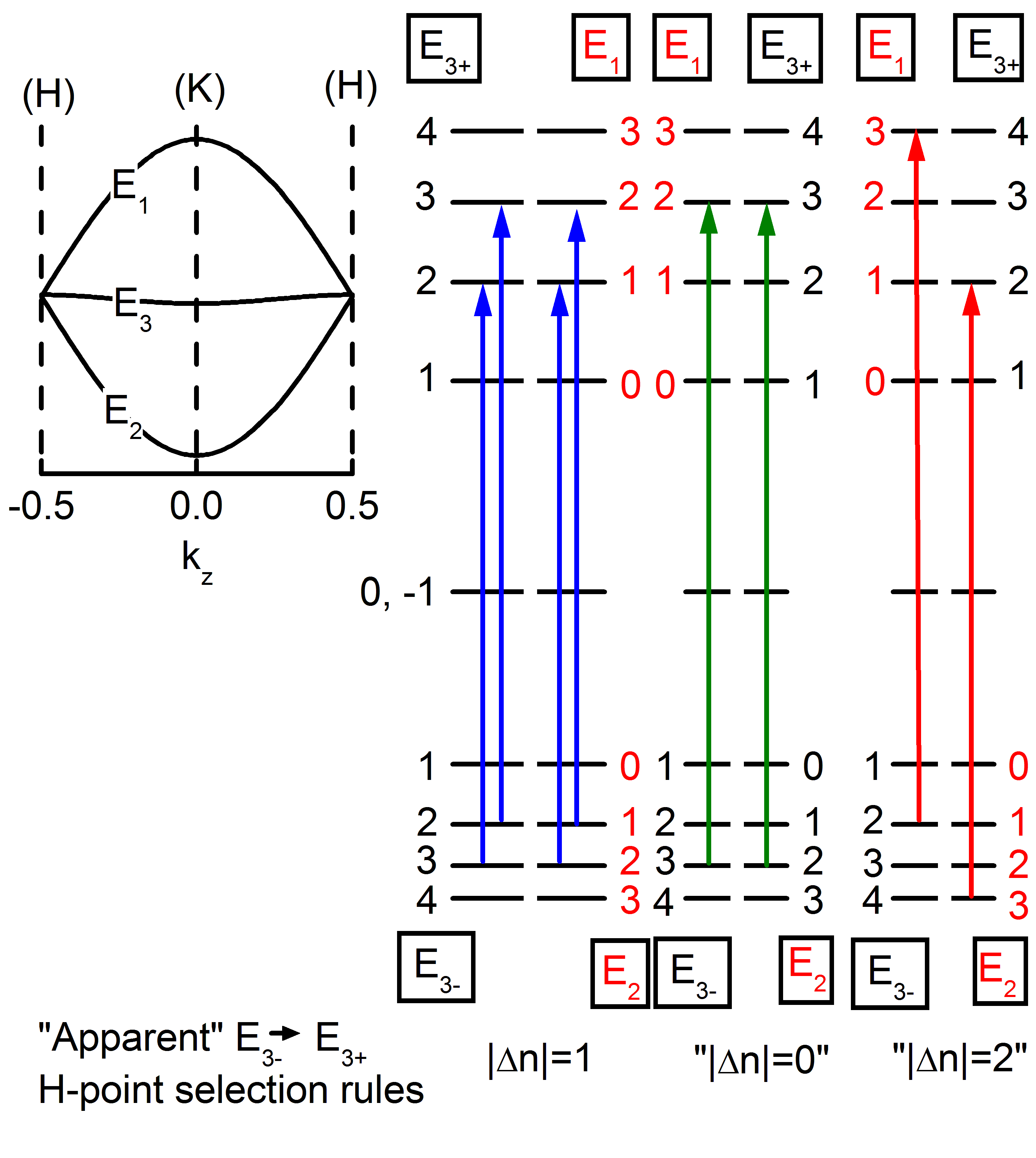}
\end{center}
\caption{(Color online) (left) Band structure of graphite along the $H-K-H$ edge. (right) Schematic of the Landau level
energies at the $H$-point showing the electron-hole asymmetry. Arrows indicate dipole allowed transitions ($\Delta
n=\pm 1$). Transitions are labeled as ``effective'' $E_{3-}\rightarrow E_{3+}$ transitions with ``apparent'' dipole
selection rules $\Delta n=\pm 1,0,\pm 2$.} \label{fig3}
\end{figure}

Representative differential absorption spectra measured at $T\simeq 1.8$~K in magnetic fields $B=55-59$~T are shown in
Fig.\ref{fig2}(a). The spectra contains a large number of lines reflecting the large number of $K$ and $H$ point
transitions which cross in this energy region. Nevertheless, a clear splitting of the $H$-point and the $K$-point
transitions is observed (arrows). The energy of the observed transitions are plotted as a function of magnetic field in
Fig.\ref{fig2}(b). Before discussing these results it is useful to consider the possible transitions at the $H$-point.
Dipole allowed transitions have a change in the orbital quantum number of $\pm 1$. Due to the doubly degenerate Landau
level spectrum at the $H$-point with $E_{3_\pm}^{n+1}=E_{1,2}^{n}$ there are a large number of allowed transitions
between the valence band ($E_{3-}$ or $E_{2}$) and the conduction band ($E_{3+}$ or $E_{1}$). However, the
understanding of the problem is greatly facilitated by fact that all transitions involving bands $E_{2}$ or $E_{1}$ are
degenerate with $E_{3-}\rightarrow E_{3+}$ transitions with ``apparent'' selection rules $\Delta n = 0$ and $\Delta n =
\pm 2$. This is shown schematically in Fig.\ref{fig3}. The electron hole asymmetry, also shown schematically here,
splits both the $\Delta n=\pm 1$ and the $\Delta n=\pm 2$ transitions, while the $\Delta n=0$ transitions remain
unaffected. From Eq.\ref{EqHpoint1} the splitting of the $\Delta n=\pm 2$ transitions is $\delta E = \hbar^2 s/m$
\emph{i.e.} twice the size of the splitting of the $\Delta n=\pm 1$ transitions.

The energy of the observed $H$ and $K$-point transitions are plotted as a function of magnetic field in
Fig.\ref{fig2}(b). As seen in the raw data, a splitting of the $H$-point and the $K$-point transitions is observed. The
calculated SWM transitions energies are indicated by the solid and broken lines.

The energy of the $H$-point transitions depends only on $\gamma_0=3.15$~eV and the calculated splitting is independent
of all other SWM parameters and vanishes only if the free electron terms are not included in the Hamiltonian. We have
verified that the predictions of Eq.(\ref{EqHpoint1}) are exact. The observed splitting of the $H$-point
$E_{3-}^{n(n+1)}\rightarrow E_{3+}^{n+1(n)}$ transitions (blue solid lines) is beautifully reproduced by the
calculations. We stress that in either approach there are no fitting parameters; the size of the splitting is simply
given by $\hbar^2 s / 2 m \simeq 0.23$~meV/T. In addition, the observed splitting of the ``effective''
$E_{3-}\rightarrow E_{3+}$ transitions with ``apparent'' selection rules $\Delta n = \pm 2$ (dashed red lines) is twice
as large in agreement with the predictions for electron hole asymmetry in Eq(\ref{Bilayer}).

\begin{table}[b]
\begin{center}
\begin{tabular}{ccc}
\hline\hline
$\gamma_0=3.15$~eV & $\gamma_1=0.37$~eV & $\gamma_2=-0.0243$~eV\\
$\gamma_3=0.31$~eV & $\gamma_4=0.07$~eV & $\gamma_5=0.05$~eV\\
$\Delta=-0.002$\\
\end{tabular}
\end{center}
\caption{Summary of the SWM parameters used.}\label{tab1}
\end{table}

The calculated splitting of the $K$-point transitions depends on the SWM parameters used, notably $\gamma_4$ and
$\gamma_5$. We adjust very slightly $\gamma_1=0.37$~eV to fit the observed transitions (slope of the magnetic field
dependence) and use the accepted values for the other SWM parameters which are summarized in Table~\ref{tab1}. The
agreement turns out to be very good making a further refinement of the parameters unnecessary. A comparison of the SWM
splitting $\simeq 23$~meV at $B=60$~T with $\hbar s/2m \simeq 14$~meV suggests that $\gamma_4$ and $\gamma_5$ are
responsible for approximately $40$\% of the splitting. The relative importance of the contribution of the free electron
kinetic energy terms to the electron-hole asymmetry means that any data analysis which neglects them would lead to a
significant over estimation of size of $\gamma_4$ or $\gamma_5$.

\begin{figure}[]
\begin{center}
\includegraphics[width= 7.5cm]{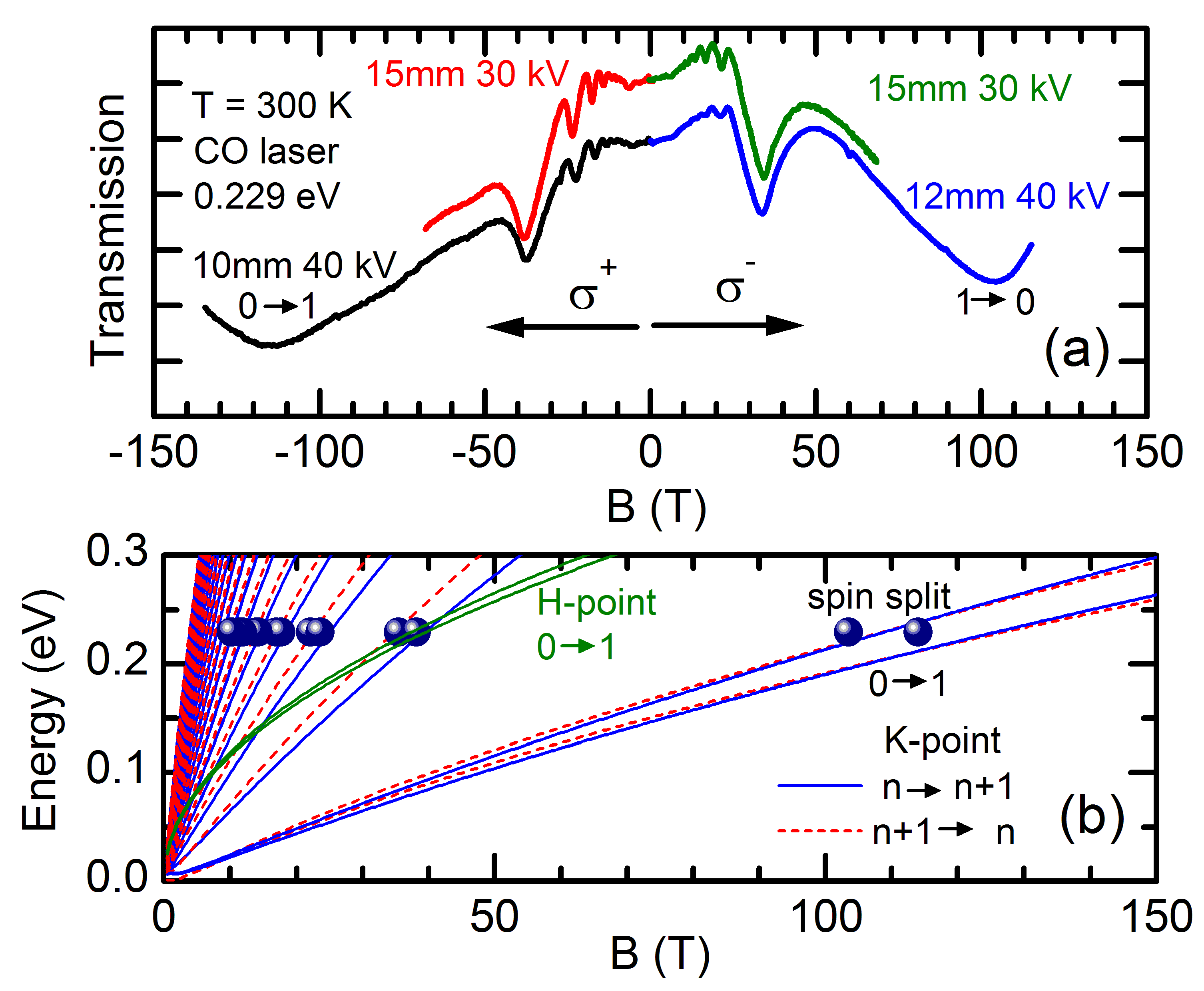}
\end{center}
\caption{(Color online) (a) Magneto-transmission of natural graphite showing mainly $K$-point transitions. (b)
Calculated SWM transitions together with the measured splitting (symbols). There is no electron-hole asymmetry for the
$0 \rightarrow 1$ K-point transition which splits due to the Zeeman term.} \label{fig4}
\end{figure}

Polarization resolved magneto-transmission, in fields up to $\pm 140$~T are shown in Fig.\ref{fig4}(a). Mainly
$K$-point transitions are observed in this energy range. The different field directions corresponds to different
polarizations and the features are shifted in field due to the different energy of the $n \rightarrow n+1$ and $n+1
\rightarrow n$ transitions. The feature around $100$~T is the fundamental $0 \rightarrow 1$ transition, which should
not be split (shifted) since the LL0 is special and has a free electron term $\simeq(n+3/2)\hbar^2 s/2m$ with $n=0$
which is identical to the free electron term of the $n=1$ Landau level $(n+1/2)\hbar^2 s/2m$. Nevertheless, the
position is shifted by $\simeq 10$~T between the two polarizations. However, this can be understood when spin splitting
is taken into account. The SWM prediction for the transitions are shown in Fig.~\ref{fig4}(b) together with the
measured field splitting. It can be seen that there is indeed no effect of electron-hole asymmetry for the $0
\rightarrow 1$ $K$-point transition (solid and broken lines). However, including the Zeeman term $\pm g \mu_B B/2$ with
$g=2$ the calculated splitting of the $0 \rightarrow 1$ transition, is comparable with the observed splitting. The
Zeeman term is important here due to the very high magnetic field, and the fact that the $0 \rightarrow 1$ transition
evolves very slowly with magnetic field so that a small energy splitting can generate a large field splitting in the
data.

Finally, we note that the Landau level energy spectrum of graphene can be derived from the SWM Hamiltonian simply by
setting all the inter-layer coupling parameters $\gamma_1,..,\gamma_5=0$. The analytic solution of this simplified
Hamiltonian is nothing other than Eq.(\ref{EqHpoint1}). This implies that the electron-hole asymmetry observed in the
cyclotron resonance of exfoliated graphene~\cite{Deacon07} also originates from the neglected free electron kinetic
energy terms.


\begin{thebibliography}{26}%
\makeatletter
\providecommand \@ifxundefined [1]{%
 \@ifx{#1\undefined}
}%
\providecommand \@ifnum [1]{%
 \ifnum #1\expandafter \@firstoftwo
 \else \expandafter \@secondoftwo
 \fi
}%
\providecommand \@ifx [1]{%
 \ifx #1\expandafter \@firstoftwo
 \else \expandafter \@secondoftwo
 \fi
}%
\providecommand \natexlab [1]{#1}%
\providecommand \enquote  [1]{``#1''}%
\providecommand \bibnamefont  [1]{#1}%
\providecommand \bibfnamefont [1]{#1}%
\providecommand \citenamefont [1]{#1}%
\providecommand \href@noop [0]{\@secondoftwo}%
\providecommand \href [0]{\begingroup \@sanitize@url \@href}%
\providecommand \@href[1]{\@@startlink{#1}\@@href}%
\providecommand \@@href[1]{\endgroup#1\@@endlink}%
\providecommand \@sanitize@url [0]{\catcode `\\12\catcode `\$12\catcode
  `\&12\catcode `\#12\catcode `\^12\catcode `\_12\catcode `\%12\relax}%
\providecommand \@@startlink[1]{}%
\providecommand \@@endlink[0]{}%
\providecommand \url  [0]{\begingroup\@sanitize@url \@url }%
\providecommand \@url [1]{\endgroup\@href {#1}{\urlprefix }}%
\providecommand \urlprefix  [0]{URL }%
\providecommand \Eprint [0]{\href }%
\providecommand \doibase [0]{http://dx.doi.org/}%
\providecommand \selectlanguage [0]{\@gobble}%
\providecommand \bibinfo  [0]{\@secondoftwo}%
\providecommand \bibfield  [0]{\@secondoftwo}%
\providecommand \translation [1]{[#1]}%
\providecommand \BibitemOpen [0]{}%
\providecommand \bibitemStop [0]{}%
\providecommand \bibitemNoStop [0]{.\EOS\space}%
\providecommand \EOS [0]{\spacefactor3000\relax}%
\providecommand \BibitemShut  [1]{\csname bibitem#1\endcsname}%
\let\auto@bib@innerbib\@empty
\bibitem [{\citenamefont {Slonczewski}\ and\ \citenamefont
  {Weiss}(1958)}]{Slonczewski58}%
  \BibitemOpen
  \bibfield  {author} {\bibinfo {author} {\bibfnamefont {J.~C.}\ \bibnamefont
  {Slonczewski}}\ and\ \bibinfo {author} {\bibfnamefont {P.~R.}\ \bibnamefont
  {Weiss}},\ }\href@noop {} {\bibfield  {journal} {\bibinfo  {journal} {Phys.
  Rev.}\ }\textbf {\bibinfo {volume} {109}},\ \bibinfo {pages} {272} (\bibinfo
  {year} {1958})}\BibitemShut {NoStop}%
\bibitem [{\citenamefont {McClure}(1960)}]{McClure60}%
  \BibitemOpen
  \bibfield  {author} {\bibinfo {author} {\bibfnamefont {J.~W.}\ \bibnamefont
  {McClure}},\ }\href@noop {} {\bibfield  {journal} {\bibinfo  {journal} {Phys.
  Rev.}\ }\textbf {\bibinfo {volume} {119}},\ \bibinfo {pages} {606} (\bibinfo
  {year} {1960})}\BibitemShut {NoStop}%
\bibitem [{\citenamefont {Nakao}(1976)}]{Nakao76}%
  \BibitemOpen
  \bibfield  {author} {\bibinfo {author} {\bibfnamefont {K.}~\bibnamefont
  {Nakao}},\ }\href@noop {} {\bibfield  {journal} {\bibinfo  {journal} {J.
  Phys. Soc. Japan}\ }\textbf {\bibinfo {volume} {40}},\ \bibinfo {pages} {761}
  (\bibinfo {year} {1976})}\BibitemShut {NoStop}%
\bibitem [{\citenamefont {Soule}\ \emph {et~al.}(1964)\citenamefont {Soule},
  \citenamefont {McClure},\ and\ \citenamefont {Smith}}]{Soule64}%
  \BibitemOpen
  \bibfield  {author} {\bibinfo {author} {\bibfnamefont {D.~E.}\ \bibnamefont
  {Soule}}, \bibinfo {author} {\bibfnamefont {J.~W.}\ \bibnamefont {McClure}},
  \ and\ \bibinfo {author} {\bibfnamefont {L.~B.}\ \bibnamefont {Smith}},\
  }\href@noop {} {\bibfield  {journal} {\bibinfo  {journal} {Phys. Rev.}\
  }\textbf {\bibinfo {volume} {134}},\ \bibinfo {pages} {A453} (\bibinfo {year}
  {1964})}\BibitemShut {NoStop}%
\bibitem [{\citenamefont {Williamson}\ \emph {et~al.}(1965)\citenamefont
  {Williamson}, \citenamefont {Foner},\ and\ \citenamefont
  {Dresselhaus}}]{Williamson65}%
  \BibitemOpen
  \bibfield  {author} {\bibinfo {author} {\bibfnamefont {S.~J.}\ \bibnamefont
  {Williamson}}, \bibinfo {author} {\bibfnamefont {S.}~\bibnamefont {Foner}}, \
  and\ \bibinfo {author} {\bibfnamefont {M.~S.}\ \bibnamefont {Dresselhaus}},\
  }\href {\doibase 10.1103/PhysRev.140.A1429} {\bibfield  {journal} {\bibinfo
  {journal} {Phys. Rev.}\ }\textbf {\bibinfo {volume} {140}},\ \bibinfo {pages}
  {A1429} (\bibinfo {year} {1965})}\BibitemShut {NoStop}%
\bibitem [{\citenamefont {Schroeder}\ \emph {et~al.}(1968)\citenamefont
  {Schroeder}, \citenamefont {Dresselhaus},\ and\ \citenamefont
  {Javan}}]{Schroeder68}%
  \BibitemOpen
  \bibfield  {author} {\bibinfo {author} {\bibfnamefont {P.~R.}\ \bibnamefont
  {Schroeder}}, \bibinfo {author} {\bibfnamefont {M.~S.}\ \bibnamefont
  {Dresselhaus}}, \ and\ \bibinfo {author} {\bibfnamefont {A.}~\bibnamefont
  {Javan}},\ }\href@noop {} {\bibfield  {journal} {\bibinfo  {journal} {Phys.
  Rev. Lett.}\ }\textbf {\bibinfo {volume} {20}},\ \bibinfo {pages} {1292}
  (\bibinfo {year} {1968})}\BibitemShut {NoStop}%
\bibitem [{\citenamefont {Woollam}(1970)}]{Woollam70}%
  \BibitemOpen
  \bibfield  {author} {\bibinfo {author} {\bibfnamefont {J.~A.}\ \bibnamefont
  {Woollam}},\ }\href@noop {} {\bibfield  {journal} {\bibinfo  {journal} {Phys.
  Rev. Lett.}\ }\textbf {\bibinfo {volume} {70}},\ \bibinfo {pages} {811}
  (\bibinfo {year} {1970})}\BibitemShut {NoStop}%
\bibitem [{\citenamefont {Woollam}(1971)}]{Woollam71a}%
  \BibitemOpen
  \bibfield  {author} {\bibinfo {author} {\bibfnamefont {J.~A.}\ \bibnamefont
  {Woollam}},\ }\href@noop {} {\bibfield  {journal} {\bibinfo  {journal} {Phys.
  Rev. B}\ }\textbf {\bibinfo {volume} {4}},\ \bibinfo {pages} {3393} (\bibinfo
  {year} {1971})}\BibitemShut {NoStop}%
\bibitem [{\citenamefont {Shimamoto}\ \emph {et~al.}(1998)\citenamefont
  {Shimamoto}, \citenamefont {Miura},\ and\ \citenamefont
  {Nojiri}}]{Shimamoto98}%
  \BibitemOpen
  \bibfield  {author} {\bibinfo {author} {\bibfnamefont {Y.}~\bibnamefont
  {Shimamoto}}, \bibinfo {author} {\bibfnamefont {N.}~\bibnamefont {Miura}}, \
  and\ \bibinfo {author} {\bibfnamefont {H.}~\bibnamefont {Nojiri}},\ }\href
  {http://stacks.iop.org/0953-8984/10/i=49/a=018} {\bibfield  {journal}
  {\bibinfo  {journal} {Journal of Physics: Condensed Matter}\ }\textbf
  {\bibinfo {volume} {10}},\ \bibinfo {pages} {11289} (\bibinfo {year}
  {1998})}\BibitemShut {NoStop}%
\bibitem [{\citenamefont {Luk'yanchuk}\ and\ \citenamefont
  {Kopelevich}(2004)}]{Lukyanchuk04}%
  \BibitemOpen
  \bibfield  {author} {\bibinfo {author} {\bibfnamefont {I.~A.}\ \bibnamefont
  {Luk'yanchuk}}\ and\ \bibinfo {author} {\bibfnamefont {Y.}~\bibnamefont
  {Kopelevich}},\ }\href@noop {} {\bibfield  {journal} {\bibinfo  {journal}
  {Phys. Rev. Lett.}\ }\textbf {\bibinfo {volume} {93}},\ \bibinfo {pages}
  {166402} (\bibinfo {year} {2004})}\BibitemShut {NoStop}%
\bibitem [{\citenamefont {Luk'yanchuk}\ and\ \citenamefont
  {Kopelevich}(2006)}]{Lukyanchuk06}%
  \BibitemOpen
  \bibfield  {author} {\bibinfo {author} {\bibfnamefont {I.~A.}\ \bibnamefont
  {Luk'yanchuk}}\ and\ \bibinfo {author} {\bibfnamefont {Y.}~\bibnamefont
  {Kopelevich}},\ }\href@noop {} {\bibfield  {journal} {\bibinfo  {journal}
  {Phys. Rev. Lett.}\ }\textbf {\bibinfo {volume} {97}},\ \bibinfo {pages}
  {256801} (\bibinfo {year} {2006})}\BibitemShut {NoStop}%
\bibitem [{\citenamefont {Mikitik}\ and\ \citenamefont {{Yu. V.
  Sharlai}}(2006)}]{Mikitik06}%
  \BibitemOpen
  \bibfield  {author} {\bibinfo {author} {\bibfnamefont {G.~P.}\ \bibnamefont
  {Mikitik}}\ and\ \bibinfo {author} {\bibnamefont {{Yu. V. Sharlai}}},\
  }\href@noop {} {\bibfield  {journal} {\bibinfo  {journal} {Phys. Rev. B}\
  }\textbf {\bibinfo {volume} {73}},\ \bibinfo {pages} {235112} (\bibinfo
  {year} {2006})}\BibitemShut {NoStop}%
\bibitem [{\citenamefont {Schneider}\ \emph {et~al.}(2009)\citenamefont
  {Schneider}, \citenamefont {Orlita}, \citenamefont {Potemski},\ and\
  \citenamefont {Maude}}]{Schneider09}%
  \BibitemOpen
  \bibfield  {author} {\bibinfo {author} {\bibfnamefont {J.~M.}\ \bibnamefont
  {Schneider}}, \bibinfo {author} {\bibfnamefont {M.}~\bibnamefont {Orlita}},
  \bibinfo {author} {\bibfnamefont {M.}~\bibnamefont {Potemski}}, \ and\
  \bibinfo {author} {\bibfnamefont {D.~K.}\ \bibnamefont {Maude}},\ }\href@noop
  {} {\bibfield  {journal} {\bibinfo  {journal} {Phys. Rev. Lett.}\ }\textbf
  {\bibinfo {volume} {102}},\ \bibinfo {pages} {166403} (\bibinfo {year}
  {2009})}\BibitemShut {NoStop}%
\bibitem [{\citenamefont {Zhu}\ \emph {et~al.}(2009)\citenamefont {Zhu},
  \citenamefont {Yang}, \citenamefont {Fauqu\'e}, \citenamefont {Kopelevich},\
  and\ \citenamefont {Behnia}}]{Zhu09}%
  \BibitemOpen
  \bibfield  {author} {\bibinfo {author} {\bibfnamefont {Z.}~\bibnamefont
  {Zhu}}, \bibinfo {author} {\bibfnamefont {H.}~\bibnamefont {Yang}}, \bibinfo
  {author} {\bibfnamefont {B.}~\bibnamefont {Fauqu\'e}}, \bibinfo {author}
  {\bibfnamefont {Y.}~\bibnamefont {Kopelevich}}, \ and\ \bibinfo {author}
  {\bibfnamefont {K.}~\bibnamefont {Behnia}},\ }\href@noop {} {\bibfield
  {journal} {\bibinfo  {journal} {Nature Physics}\ }\textbf {\bibinfo {volume}
  {6}},\ \bibinfo {pages} {26} (\bibinfo {year} {2009})}\BibitemShut {NoStop}%
\bibitem [{\citenamefont {Schneider}\ \emph
  {et~al.}(2010{\natexlab{a}})\citenamefont {Schneider}, \citenamefont
  {Goncharuk}, \citenamefont {Vasek}, \citenamefont {Svoboda}, \citenamefont
  {Vyborny}, \citenamefont {Smrcka}, \citenamefont {Orlita}, \citenamefont
  {Potemski},\ and\ \citenamefont {Maude}}]{Schneider10}%
  \BibitemOpen
  \bibfield  {author} {\bibinfo {author} {\bibfnamefont {J.~M.}\ \bibnamefont
  {Schneider}}, \bibinfo {author} {\bibfnamefont {N.~A.}\ \bibnamefont
  {Goncharuk}}, \bibinfo {author} {\bibfnamefont {P.}~\bibnamefont {Vasek}},
  \bibinfo {author} {\bibfnamefont {P.}~\bibnamefont {Svoboda}}, \bibinfo
  {author} {\bibfnamefont {Z.}~\bibnamefont {Vyborny}}, \bibinfo {author}
  {\bibfnamefont {L.}~\bibnamefont {Smrcka}}, \bibinfo {author} {\bibfnamefont
  {M.}~\bibnamefont {Orlita}}, \bibinfo {author} {\bibfnamefont
  {M.}~\bibnamefont {Potemski}}, \ and\ \bibinfo {author} {\bibfnamefont
  {D.~K.}\ \bibnamefont {Maude}},\ }\href {\doibase 10.1103/PhysRevB.81.195204}
  {\bibfield  {journal} {\bibinfo  {journal} {Phys. Rev. B}\ }\textbf {\bibinfo
  {volume} {81}},\ \bibinfo {pages} {195204} (\bibinfo {year}
  {2010}{\natexlab{a}})}\BibitemShut {NoStop}%
\bibitem [{\citenamefont {Luk'yanchuk}\ and\ \citenamefont
  {Kopelevich}(2010)}]{Lukyanchuk10}%
  \BibitemOpen
  \bibfield  {author} {\bibinfo {author} {\bibfnamefont {I.~A.}\ \bibnamefont
  {Luk'yanchuk}}\ and\ \bibinfo {author} {\bibfnamefont {Y.}~\bibnamefont
  {Kopelevich}},\ }\href@noop {} {\bibfield  {journal} {\bibinfo  {journal}
  {Phys. Rev. Lett.}\ }\textbf {\bibinfo {volume} {104}},\ \bibinfo {pages}
  {119701} (\bibinfo {year} {2010})}\BibitemShut {NoStop}%
\bibitem [{\citenamefont {Schneider}\ \emph
  {et~al.}(2010{\natexlab{b}})\citenamefont {Schneider}, \citenamefont
  {Orlita}, \citenamefont {Potemski},\ and\ \citenamefont
  {Maude}}]{Schneider10a}%
  \BibitemOpen
  \bibfield  {author} {\bibinfo {author} {\bibfnamefont {J.~M.}\ \bibnamefont
  {Schneider}}, \bibinfo {author} {\bibfnamefont {M.}~\bibnamefont {Orlita}},
  \bibinfo {author} {\bibfnamefont {M.}~\bibnamefont {Potemski}}, \ and\
  \bibinfo {author} {\bibfnamefont {D.~K.}\ \bibnamefont {Maude}},\ }\href@noop
  {} {\bibfield  {journal} {\bibinfo  {journal} {Phys. Rev. Lett.}\ }\textbf
  {\bibinfo {volume} {104}},\ \bibinfo {pages} {119702} (\bibinfo {year}
  {2010}{\natexlab{b}})}\BibitemShut {NoStop}%
\bibitem [{\citenamefont {Hubbard}\ \emph {et~al.}(2011)\citenamefont
  {Hubbard}, \citenamefont {Kershaw}, \citenamefont {Usher}, \citenamefont
  {Savchenko},\ and\ \citenamefont {Shytov}}]{Hubbard11}%
  \BibitemOpen
  \bibfield  {author} {\bibinfo {author} {\bibfnamefont {S.~B.}\ \bibnamefont
  {Hubbard}}, \bibinfo {author} {\bibfnamefont {T.~J.}\ \bibnamefont
  {Kershaw}}, \bibinfo {author} {\bibfnamefont {A.}~\bibnamefont {Usher}},
  \bibinfo {author} {\bibfnamefont {A.~K.}\ \bibnamefont {Savchenko}}, \ and\
  \bibinfo {author} {\bibfnamefont {A.}~\bibnamefont {Shytov}},\ }\href
  {\doibase 10.1103/PhysRevB.83.035122} {\bibfield  {journal} {\bibinfo
  {journal} {Phys. Rev. B}\ }\textbf {\bibinfo {volume} {83}},\ \bibinfo
  {pages} {035122} (\bibinfo {year} {2011})}\BibitemShut {NoStop}%
\bibitem [{\citenamefont {Doezema}\ \emph {et~al.}(1979)\citenamefont
  {Doezema}, \citenamefont {Datars}, \citenamefont {Schaber},\ and\
  \citenamefont {Van~Schyndel}}]{Doezema79}%
  \BibitemOpen
  \bibfield  {author} {\bibinfo {author} {\bibfnamefont {R.~E.}\ \bibnamefont
  {Doezema}}, \bibinfo {author} {\bibfnamefont {W.~R.}\ \bibnamefont {Datars}},
  \bibinfo {author} {\bibfnamefont {H.}~\bibnamefont {Schaber}}, \ and\
  \bibinfo {author} {\bibfnamefont {A.}~\bibnamefont {Van~Schyndel}},\ }\href
  {\doibase 10.1103/PhysRevB.19.4224} {\bibfield  {journal} {\bibinfo
  {journal} {Phys. Rev. B}\ }\textbf {\bibinfo {volume} {19}},\ \bibinfo
  {pages} {4224} (\bibinfo {year} {1979})}\BibitemShut {NoStop}%
\bibitem [{\citenamefont {Orlita}\ \emph {et~al.}(2008)\citenamefont {Orlita},
  \citenamefont {Faugeras}, \citenamefont {Martinez}, \citenamefont {Maude},
  \citenamefont {Sadowski},\ and\ \citenamefont {Potemski}}]{Orlita08}%
  \BibitemOpen
  \bibfield  {author} {\bibinfo {author} {\bibfnamefont {M.}~\bibnamefont
  {Orlita}}, \bibinfo {author} {\bibfnamefont {C.}~\bibnamefont {Faugeras}},
  \bibinfo {author} {\bibfnamefont {G.}~\bibnamefont {Martinez}}, \bibinfo
  {author} {\bibfnamefont {D.~K.}\ \bibnamefont {Maude}}, \bibinfo {author}
  {\bibfnamefont {M.~L.}\ \bibnamefont {Sadowski}}, \ and\ \bibinfo {author}
  {\bibfnamefont {M.}~\bibnamefont {Potemski}},\ }\href {\doibase
  10.1103/PhysRevLett.100.136403} {\bibfield  {journal} {\bibinfo  {journal}
  {Phys. Rev. Lett.}\ }\textbf {\bibinfo {volume} {100}},\ \bibinfo {pages}
  {136403} (\bibinfo {year} {2008})}\BibitemShut {NoStop}%
\bibitem [{\citenamefont {Orlita}\ \emph {et~al.}(2009)\citenamefont {Orlita},
  \citenamefont {Faugeras}, \citenamefont {Schneider}, \citenamefont
  {Martinez}, \citenamefont {Maude},\ and\ \citenamefont
  {Potemski}}]{Orlita09}%
  \BibitemOpen
  \bibfield  {author} {\bibinfo {author} {\bibfnamefont {M.}~\bibnamefont
  {Orlita}}, \bibinfo {author} {\bibfnamefont {C.}~\bibnamefont {Faugeras}},
  \bibinfo {author} {\bibfnamefont {J.~M.}\ \bibnamefont {Schneider}}, \bibinfo
  {author} {\bibfnamefont {G.}~\bibnamefont {Martinez}}, \bibinfo {author}
  {\bibfnamefont {D.~K.}\ \bibnamefont {Maude}}, \ and\ \bibinfo {author}
  {\bibfnamefont {M.}~\bibnamefont {Potemski}},\ }\href {\doibase
  10.1103/PhysRevLett.102.166401} {\bibfield  {journal} {\bibinfo  {journal}
  {Phys. Rev. Lett.}\ }\textbf {\bibinfo {volume} {102}},\ \bibinfo {pages}
  {166401} (\bibinfo {year} {2009})}\BibitemShut {NoStop}%
\bibitem [{\citenamefont {Chuang}\ \emph {et~al.}(2009)\citenamefont {Chuang},
  \citenamefont {Baker},\ and\ \citenamefont {Nicholas}}]{Chuang09}%
  \BibitemOpen
  \bibfield  {author} {\bibinfo {author} {\bibfnamefont {K.-C.}\ \bibnamefont
  {Chuang}}, \bibinfo {author} {\bibfnamefont {A.~M.~R.}\ \bibnamefont
  {Baker}}, \ and\ \bibinfo {author} {\bibfnamefont {R.~J.}\ \bibnamefont
  {Nicholas}},\ }\href {\doibase 10.1103/PhysRevB.80.161410} {\bibfield
  {journal} {\bibinfo  {journal} {Phys. Rev. B}\ }\textbf {\bibinfo {volume}
  {80}},\ \bibinfo {pages} {161410} (\bibinfo {year} {2009})}\BibitemShut
  {NoStop}%
\bibitem [{\citenamefont {Ubrig}\ \emph {et~al.}(2011)\citenamefont {Ubrig},
  \citenamefont {Plochocka}, \citenamefont {Kossacki}, \citenamefont {Orlita},
  \citenamefont {Maude}, \citenamefont {Portugall},\ and\ \citenamefont
  {Rikken}}]{Ubrig11}%
  \BibitemOpen
  \bibfield  {author} {\bibinfo {author} {\bibfnamefont {N.}~\bibnamefont
  {Ubrig}}, \bibinfo {author} {\bibfnamefont {P.}~\bibnamefont {Plochocka}},
  \bibinfo {author} {\bibfnamefont {P.}~\bibnamefont {Kossacki}}, \bibinfo
  {author} {\bibfnamefont {M.}~\bibnamefont {Orlita}}, \bibinfo {author}
  {\bibfnamefont {D.~K.}\ \bibnamefont {Maude}}, \bibinfo {author}
  {\bibfnamefont {O.}~\bibnamefont {Portugall}}, \ and\ \bibinfo {author}
  {\bibfnamefont {G.~L. J.~A.}\ \bibnamefont {Rikken}},\ }\href {\doibase
  10.1103/PhysRevB.83.073401} {\bibfield  {journal} {\bibinfo  {journal} {Phys.
  Rev. B}\ }\textbf {\bibinfo {volume} {83}},\ \bibinfo {pages} {073401}
  (\bibinfo {year} {2011})}\BibitemShut {NoStop}%
\bibitem [{\citenamefont {Tung}\ \emph {et~al.}(2011)\citenamefont {Tung},
  \citenamefont {Cadden-Zimansky}, \citenamefont {Qi}, \citenamefont {Jiang},\
  and\ \citenamefont {Smirnov}}]{Tung11}%
  \BibitemOpen
  \bibfield  {author} {\bibinfo {author} {\bibfnamefont {L.-C.}\ \bibnamefont
  {Tung}}, \bibinfo {author} {\bibfnamefont {P.}~\bibnamefont
  {Cadden-Zimansky}}, \bibinfo {author} {\bibfnamefont {J.}~\bibnamefont {Qi}},
  \bibinfo {author} {\bibfnamefont {Z.}~\bibnamefont {Jiang}}, \ and\ \bibinfo
  {author} {\bibfnamefont {D.}~\bibnamefont {Smirnov}},\ }\href {\doibase
  10.1103/PhysRevB.84.153405} {\bibfield  {journal} {\bibinfo  {journal} {Phys.
  Rev. B}\ }\textbf {\bibinfo {volume} {84}},\ \bibinfo {pages} {153405}
  (\bibinfo {year} {2011})}\BibitemShut {NoStop}%
\bibitem [{\citenamefont {Koshino}\ and\ \citenamefont
  {Ando}(2008)}]{Koshino08}%
  \BibitemOpen
  \bibfield  {author} {\bibinfo {author} {\bibfnamefont {M.}~\bibnamefont
  {Koshino}}\ and\ \bibinfo {author} {\bibfnamefont {T.}~\bibnamefont {Ando}},\
  }\href {\doibase 10.1103/PhysRevB.77.115313} {\bibfield  {journal} {\bibinfo
  {journal} {Phys. Rev. B}\ }\textbf {\bibinfo {volume} {77}},\ \bibinfo
  {pages} {115313} (\bibinfo {year} {2008})}\BibitemShut {NoStop}%
\bibitem [{\citenamefont {Deacon}\ \emph {et~al.}(2007)\citenamefont {Deacon},
  \citenamefont {Chuang}, \citenamefont {Nicholas}, \citenamefont {Novoselov},\
  and\ \citenamefont {Geim}}]{Deacon07}%
  \BibitemOpen
  \bibfield  {author} {\bibinfo {author} {\bibfnamefont {R.~S.}\ \bibnamefont
  {Deacon}}, \bibinfo {author} {\bibfnamefont {K.-C.}\ \bibnamefont {Chuang}},
  \bibinfo {author} {\bibfnamefont {R.~J.}\ \bibnamefont {Nicholas}}, \bibinfo
  {author} {\bibfnamefont {K.~S.}\ \bibnamefont {Novoselov}}, \ and\ \bibinfo
  {author} {\bibfnamefont {A.~K.}\ \bibnamefont {Geim}},\ }\href {\doibase
  10.1103/PhysRevB.76.081406} {\bibfield  {journal} {\bibinfo  {journal} {Phys.
  Rev. B}\ }\textbf {\bibinfo {volume} {76}},\ \bibinfo {pages} {081406}
  (\bibinfo {year} {2007})}\BibitemShut {NoStop}%
\end{thebibliography}

%

\end{document}